# Einstein's Equivalence Principle and the Gravitational Red Shift II

By Petros S. Florides

School of Mathematics, Trinity College, Dublin 2, Ireland

E-mail: florides@maths.tcd.ie

Long before the general theory of relativity was finally formulated in 1916, arguments based entirely on Einstein's equivalence principle predicted the well known phenomenon of the gravitational red shift. Precisely the same arguments are widely being used today to derive the same phenomenon. Accordingly, it is often claimed that the observed gravitational red shift is a verification of the equivalence principle rather than a verification of the full theory of general relativity. Here we show that, contrary to these claims, the arguments based on the equivalence principle are false and that *only the full theory of general relativity can correctly and unambiguously predict the gravitational red shift.*

1. Introduction

No sooner had Einstein[1] published his special theory of relativity in 1905 than the search for a generalisation of this theory began. The search for a theory which, unlike the special theory of relativity, could deal satisfactorily with gravitational phenomena, and in which absolute motion of any kind and the existence of any preferential reference systems disappear completely; a theory which would, so it was hoped, incorporate the so-called Mach's principle.

It was, as is well known, after ten years of hard and intense work that Einstein's efforts were finally crowned with brilliant success in 1916. The outcome was his general theory of relativity[2], a beautiful geometric theory which fused not only space and time, as special relativity did, but space, time, matter and gravitation.

Throughout these ten long years "of searching in the dark," as Einstein put it, the only physical clue that guided Einstein was the experimentally well established equality between the inertial and the gravitational mass. Using this equality or, equivalently, Galileo's law that "*all bodies fall equally fast in a gravitational field*", and a series of thought experiments performed in the by now famous "*Einstein elevator*," he was able to conclude, and postulate in the form of a principle, that

> **gravitational effects on any physical process are locally completely equivalent to inertial effects.**

This is the famous *Einstein's equivalence principle* ( or EP for short), often referred to as the *strong* EP. It is remarkable that Einstein[3] was able to announce this principle, in almost exactly the above form, as early as 1907. To the overwhelming majority of relativists the EP forms the cornerstone of the general theory of relativity.

The heuristic value of the $EP$, both from the physical and from the theoretical points of view, cannot be overemphasised. For, as is well known, the EP ( together with the principle of *general covariance* and the principle of *minimum gravitational coupling* ) leads, almost with inevitable necessity, to the general theory of relativity. From the physical point of view the usefulness of the EP is that, as Einstein[3] had already pointed out in 1907, we can determine how various physical processes take place in an accelerated reference frame; by invoking the EP we can then conclude that the same physical processes take place in exactly the same way, at least locally, in a corresponding gravitational field.

By far the most important physical consequences of the EP, obtained in the way just described, are the *gravitational red shift* and the *gravitational time dilation*. Consider an emitter, $E$, e.g. a vibrating atom, at rest at a point near the Earth's surface, say, of gravitational potential $\Phi$. Let it send light, or any other electromagnetic, signals to a receiver $R$ at rest directly above $E$ and distance $h$ from it; the gravitational potential at $R$ is $\Phi + \triangle\Phi$, where $\triangle\Phi = gh$, $g$ being the acceleration due to gravity. Let $\nu_E$ be the frequency of the signal as measured at $E$, and $\nu_R$ the frequency of the signal received, and measured, at $R$. Then by an argument, to be referred to as the *standard argument* in the sequel, which is reproduced in section 2 below, it is claimed that

$$\frac{\nu_R - \nu_E}{\nu_E} = -\frac{\triangle\Phi}{c^2} = -\frac{gh}{c^2} \qquad (1)$$

Thus $\nu_E > \nu_R$, which leads us to the conclusion that light moving upwards in a static gravitational field, from a point at potential $\Phi$ to a point at higher potential $\Phi + \triangle\Phi$, becomes red-shifted. This *gravitational red shift* has been verified in terrestrial experiments by Pound and Rebka[4] in 1960 and by Pound and Snider[5] in 1964.

Suppose that $n$ waves, $n$ wave crests say, are emitted by $E$ during the small time interval $\triangle t_E$ (as measured by a clock at rest at $E$). Then, from the definition of frequency, we have $n = \nu_E \triangle t_E$. Let $R$ receive these same $n$ waves during the small time interval $\triangle t_R$ (as measured by a clock at rest at $R$). Then, again from the definition of frequency, we have $n = \nu_R \triangle t_R (= \nu_E \triangle t_E)$. Hence, correct to order $c^{-2}$, equation (1) leads to the *gravitational time dilation* formula

$$\frac{\triangle t_R}{\triangle t_E} = \frac{\nu_E}{\nu_R} = 1 + \frac{gh}{c^2}. \qquad (2)$$

Formula (2) is of paramount importance. For simple arguments, initiated by Schild[6] in 1967 and repeated in almost all the major books on relativity ( e.g., Misner[7] *et al.*, Strauman[8]), based on this formula, show that a consistent theory of gravity cannot be constructed within the framework of the special theory of relativity; in the presence of gravity the space-time manifold (of special relativity) must be replaced by a (curved) *Riemannian manifold*. To the present author this conclusion is *the single most important consequence of the EP*, in so far, of course, *as the gravitational time dilation formula (2) follows from the EP*. Alas! this is *not* the case. As we shall show below, equations (1) and (2) are in *no way consequences of the EP*.

It must be added that other arguments, based on the conservation of energy and on classical and quantum mechanics, also lead to equations (1) and (2). Since, however, these arguments use, at most, the *weak* equivalence principle (that is, the equality of inertial and gravitational mass), they shall not be considered here. To avoid any misunderstanding we stress that by the "equivalence principle" we mean the "*strong* equivalence principle" throughout the paper.

In section 2 we give the *standard argument*, based on the *EP* and *classical mechanics*, which claims to lead to equation (1) and, hence, to equation (2); it can be found in almost all the books on general relativity (e.g., Misner[7] *et al.*, Strauman[8], Adler[9] *et al.*, Sciama[10]) and is essentially the same as Einstein's argument[11] in his original Prague paper 1911. It is shown, in the same section, that the argument is fundamentally flawed in two important respects. A correct argument, based again on the *EP* and *classical mechanics*, is given in section 3 where it is shown that the *EP does not lead to any gravitational red shift*.

The stunning consequence of this result is discussed in section 4. An *abstract* of this paper appeared in the Conference Proceedings of the 2001 Journées Relativistes, which were published in a special issue of the Int. J. of Mod. Physics[12] in 2002; the Roman numeral $II$ is used in the title of the present paper to distinguish it from the aforesaid *Abstract*. A short version of the paper received Honorary Mention in 1999 as a Gravity Research Foundation Essay. The paper, in almost identical form, appeared in the series of Mathematics Reports of the School of Mathematics of Trinity College Dublin in 2011; it is available at $http://www.maths.tcd.ie/report\_series/tcdmath/tcdm1111.pdf$.

## 2. The standard argument

Consider, again, the experiment described in section 1 in which the emitter ($E$) and receiver ($R$) are at rest near the Earth's surface with $R$ distance $h$ directly above $E$. According to the EP we can, instead, consider $E$ and $R$ as fixed in an elevator which is accelerating relative to an inertial frame $S$ in gravitation-free space with constant acceleration $g$ in the direction $\vec{ER}$. At time $t = 0$, when $E$ is assumed to be at rest in $S$, $E$ emits a light wave towards $R$. The time it takes the wave to reach $R$ is roughly $t = h/c$, where $c$ is the speed of light in $S$. But in this time $R$ has acquired the velocity $V = gt = \frac{gh}{c}$. Thus, using the *exact* classical Doppler shift formula

$$\nu_{\text{receiver}} = \nu_{\text{emitter}} \left(1 - \frac{V}{c}\right), \qquad (3)$$

there is a consequent Doppler shift given by

$$\nu_R = \nu_E \left(1 - \frac{gh}{c^2}\right), \qquad (4a)$$

or

$$\frac{\nu_R - \nu_E}{\nu_E} = -\frac{gh}{c^2}. \qquad (4b)$$

By the EP the same result must hold when $E$ and $R$ are fixed near the Earth's surface. In this case $gh = \triangle\Phi = \Phi(R) - \Phi(E)$, so that in a gravitational field, near the Earth's surface in this case,

$$\frac{\nu_R - \nu_E}{\nu_E} = -\frac{gh}{c^2} = -\frac{\triangle\Phi}{c^2} \qquad (4c)$$

as in equation (1).

The above argument is, almost word by word, the same as in all the standard books on general relativity (some of which were cite in section 1). We note that, in so far as the derivation of equation(4c), and consequently of equation (2), relies on the validity of the *EP* only, the full general theory of relativity is, indeed, not needed for the derivation of the gravitational red shift (Adler[9] *et al.*).

The above *standard argument* is so simple and self evident that, for over one hundred years now, nobody (including the present author) ever questioned it. Yet, as the author discovered recently to his utter dismay,the argument is, in fact, *false*. Before we demonstrate this it is essential to make the following remarks concerning the above argument.

(i) The use of the formula $V = gt$, (more generally, $V = u + gt$), indicates clearly that classical mechanics is being used.

(ii) The elevator is assumed to be a *rigid body*, in the classical sense; that is, every point of the elevator, and in particular $E$ and $R$, have the *same acceleration* ($g$) relative to the inertial frame $S$.

We remark, further, that

(iii) the assumption that the light is emitted by $E$ at $t = 0$, when the elevator is assumed to be at rest relative to $S$, is introduced only for simplicity and should be of no consequence at all. Because acceleration is an *absolute* quantity in classical mechanics, that is *acceleration is the same in all inertial frames*, the same result must hold at whatever time the light is emitted. This, of course, *must* be the case. For, in the case when the elevator is at rest near the Earth's surface, the time at which the light is emitted is completely irrelevant due to the static character of the gravitational field. Thus, if the constantly accelerating elevator is to have any semblance to the elevator at rest on the Earth's surface, the time at which the light is emitted (in the accelerating elevator) must also be irrelevant. The importance of this remark will become clear further on.

Simple and straight forward as the standard argument may seem it is, in fact, *fundamentally wrong in the following two important respects*:

(a) In the derivation of the basic formula (3), namely

$$\nu_{\text{receiver}} = \nu_{\text{emitter}} \left(1 - \frac{V}{c}\right), \qquad (3)$$

for the *classical* Doppler shift (which, it may be recalled, is the first approximation in $V/c$ of the corresponding special relativistic formula), on which the standard argument is so decisively based, the emitter and the receiver move with *constant velocities* relative to an inertial frame and $V$ is the *constant* velocity of the receiver *relative to the emitter and away from it*. That is, the velocity of the emitter is the same at the instant of the emission of each of the $n$ waves and, likewise, the velocity of the receiver is the same at the instant of the reception of each of the $n$ waves. This is *not* the case when $E$ and $R$ are *accelerating* relative to an inertial frame. Consequently, there is no reason whatsoever why formula (3) should hold for *accelerating* emitter and receiver.

(b) In the standard argument above, the emitter (E) and receiver (R) are supposed to be moving with the same constant acceleration relative to an inertial frame. Although the standard argument does take into account the fact that $R$ is moving with constant acceleration, ( recall the use of the formula $V = u + gt$ for finding the velocity of $R$ at time $t = h/c$), it *does not, in any way, take into account the fact that $E$ is also moving with constant acceleration*. Indeed, as long as (at $t = 0$, in the standard argument) $E$ is at rest in the inertial frame $S$ and at distance $h$ from the accelerating receiver (R), the standard argument will give precisely the same result (4b) *irrespective of the general motion of $E$ before and after the instant $t = 0$*. As far as the standard argument is concerned, $E$ may just as well be *permanently* at rest in the inertial frame while $R$ moves with constant acceleration. As will be clearly demonstrated in the next section, case III, the *Doppler shift formula (4b) holds only when $E$ is permanently at rest in the inertial frame and $R$ moves with constant acceleration.* Consequently, formula (4b) has nothing to do with the above experiment in the rigid constantly accelerating elevator.

It is remarkable that these two elementary flaws in the standard argument have remained undetected for so long. In view of these flaws the standard argument falls apart. A new argument, freed from these flaws, is given in the following section.

### 3. The new argument

It is clear that any attempt to derive the gravitational red shift from the EP will necessitate a new Doppler shift formula which is valid when both $E$ and $R$ undergo constant accelerations. Below we shall derive, within the framework of *classical mechanics*, an *exact* generalised Doppler shift formula (see equation (11)) which is valid when both $E$ and $R$ undergo *arbitrary* motions along the $x$-axis of an *arbitrary* inertial frame $S$. It is evident that an *assumption* concerning the propagation of light in classical mechanics must be made. Here, as Bondi[13] pointed out in his book on cosmology, "the intrinsic weakness of Newtonian methods reveals itself. Owing to the artificial division between dynamics and electromagnetism the dynamical behaviour of the system does not determine the properties of light".

As in *Newtonian Cosmology* we shall follow McCrea and Milne[14] and Bondi[13] and state explicitly that we shall assume the validity of the *Newtonian law of the addition of velocities even if one of these velocities is the velocity of light.* For brevity we shall, in the sequel, refer to this assumption as the *Newtonian addition of velocities*. Classical mechanics will, of course, be used throughout this section as in the *standard argument*.

The justification for assuming the Newtonian addition of velocities is fourfold:

(A) It is the simplest assumption consistent with classical mechanics.

(B) On the basis of this assumption we derive (see special case **I** below) the exact Doppler shift formula (3) ( when $E$ and $R$ are moving with *constant* velocities relative to an inertial frame ) independently of the time at which emission begins and independently of the inertial frame used.

(C) Most compellingly, precisely the same assumption was made by *Bondi*[13], and others, in discussing the immensely successful theory of Newtonian cosmology of McCrea and Milne[14]. [ Recognising the importance of the McCrea-Milne paper the Journal of General Relativity and Gravitation reproduced it as one of the "Golden Oldies" papers on relativity and cosmology in its September 2000 issue]. Under this assumption the *formula for the cosmological red shift in Newtonian cosmology is identical to the corresponding formula in general relativistic cosmology.*

(D) Equally compellingly, precisely the same assuption was made by Florides and McCrea[15] when discussing the "congestion" of the universe in all the cosmological models. Again, under this assumption *the formula for the congestion in Newtonian cosmology becomes identical to that of the general relativistic cosmology.*

In view of the above results we can use the *Newtonian addition of velocities* with confidence. In any case, one cannot replace this assumption by the assumption of the constancy of the velocity of light as in special relativity unless one abandons classical mechanics altogether.

Consider an emitter ($E$) and a receiver ($R$) undergoing *arbitrary* motion along the $x$-axis, say, of the inertial frame $S$. Let their histories (or world lines) $C_E$ and $C_R$, (fig.1), be

given by the equations

$$C_E : x = x_E(t), \quad C_R : x = x_R(t) \qquad (5)$$

where $x_E(t)$ and $x_R(t)$ are assumed to be differentiable functions of $t$.

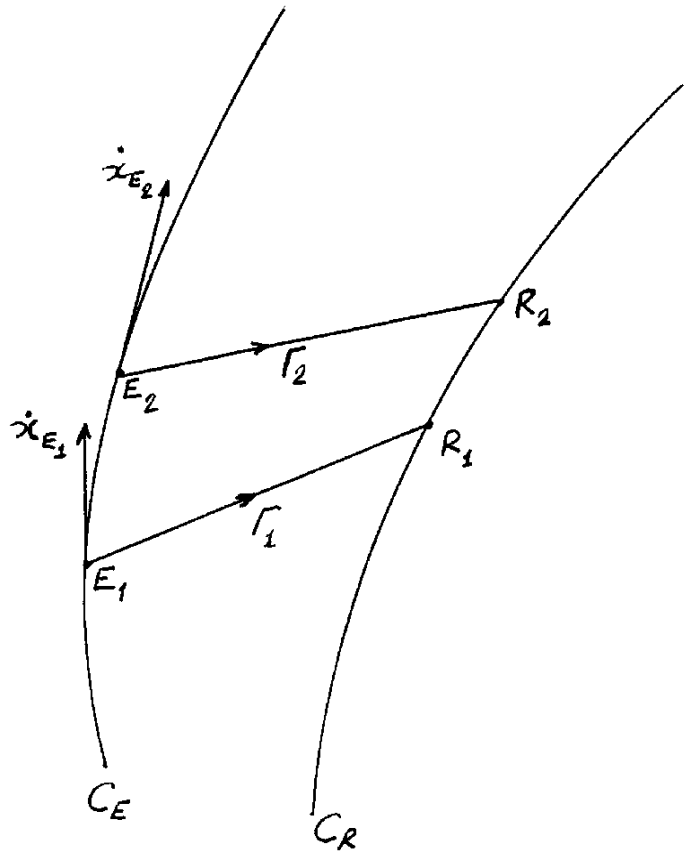

fig. 1

Let $E_1(x_1, t_1)$, with $x_1 = x_E(t_1)$, be the event on $C_E$ at which $E$ starts emitting light waves towards $R$ with velocity $c_0$, say, *relative* to $E$. Since at $t = t_1$, $E$ has velocity $\dot{x}_E(t_1)$ relative to $S$, the velocity, $c$, of the first wave relative to $S$ is, according to our assumption of the Newtonian addition of velocities,

$$c = \dot{x}_E(t_1) + c_0, \quad \left(\dot{x}_E = \frac{dx_E}{dt}\right). \qquad (6)$$

The wave retains this constant velocity in $S$ for all time. Hence the light ray (the history of the wave crest, say) is the straight line, $\Gamma_1$, whose equation is $x - x_E(t_1) = c(t - t_1)$ or, by equation (6),

$$x = x_E(t_1) + \dot{x}_E(t_1)(t - t_1) + c_0(t - t_1). \qquad (7)$$

This straight line meets the worldline, $C_R$, of the receiver $R$ at the event $R_1(x_R(T_1), T_1)$ where $T_1$ is given by

$$x_R(T_1) = x_E(t_1) + \dot{x}_E(t_1)(T_1 - t_1) + c_0(T_1 - t_1). \qquad (8)$$

Likewise let $E_2(x_2, t_2)$, with $x_2 = x_E(t_2)$, be the event on $C_E$ of the last wave leaving $E$ at time $t_2$, the velocity of this wave relative to $E$ being, again, $c_0$. The corresponding light ray, $\Gamma_2$, meets $C_R$ at the event $R_2(x_R(T_2), T_2)$ where $T_2$ is given by

$$x_R(T_2) = x_E(t_2) + \dot{x}_E(t_2)(T_2 - t_2) + c_0(T_2 - t_2). \qquad (9)$$

It is to be noted that both equations (8) and (9) are *exact*. Assume, now, that the duration of emission and the corresponding duration of reception of the waves are *small*. We can then write $t_2 = t_1 + \triangle t_E$ and $T_2 = T_1 + \triangle t_R$ where the squares of $\triangle t_E$ and $\triangle t_R$, and their product, can be neglected. Expanding equation (9) by Taylor's theorem, retaining terms linear in $\triangle t_E$ and $\triangle t_R$ only, and subtracting (8) from the resulting equation we get, after some rearrangement,

$$[\dot{x}_R(T_1) - \dot{x}_E(t_1) - c_0]\triangle t_R = [\ddot{x}_E(t_1)(T_1 - t_1) - c_0]\triangle t_E, \quad \left(\ddot{x}_E = \frac{d^2x}{dt^2}\right). \quad (10)$$

Assuming that $n$ waves are emitted during the time interval $\triangle t_E$ and that the same number of waves are received during the time interval $\triangle t_R$ then, from the definition of frequency we have, in an obvious notation, $n = \nu_E \triangle t_E = \nu_R \triangle t_R$. Using this relation in equation (10) we get the *general Doppler shift formula*

$$[c_0 - \ddot{x}_E(t_1)(T_1 - t_1)]\nu_R = [c_0 - \dot{x}_R(T_1) + \dot{x}_E(t_1)]\nu_E. \quad (11)$$

This important formula relates the frequency at reception ($\nu_R$) to the frequency at emission ($\nu_E$). The appearance of the *acceleration*, $\ddot{x}_E(t_1)$, of the emitter $E$ at time $t = t_1$, is to be *particularly* noted. Also to be noted is the fact that the acceleration of $R$ does not appear *explicitly* in equation (11); this is due to the fact that equation (11) is correct only to the first order in $\triangle t_E$ and $\triangle t_R$. We mention, in addition, that in deriving the general formula (11), it was tacitly assumed that the light rays $\Gamma_1$ and $\Gamma_2$ intersect the world line $C_R$; that is, that the solutions $T_1$ and $T_2$ given by equations (8) and (9), respectively, do exist. Whether $T_1$ and $T_2$ exist or not, and whether they are unique, depends, of course, on the functions $x_E(t)$ and $x_R(t)$.

The following three special cases of equation (11) are of crucial importance to this work:

## I. $E$ and $R$ moving with constant speeds in $S$

Let $u$ and $v$ be the speeds of $E$ and $R$, respectively, along the positive $x$-axis. We assume, for definiteness, that $v > u$ so that $R$ moves away from $E$. Then $\dot{x}_E = u$, $\ddot{x}_E = 0$, $\dot{x}_R = v$ for all $t$, and equation (11) gives the usual Doppler shift formula

( see equation (3))

$$\nu_R = \left(1 - \frac{V}{c_0}\right)\nu_E. \quad (12)$$

Here $V = v - u$ is the velocity of $R$ *relative* to $E$ and $c_0$ is the velocity of light relative to E, or relative to the *rest* inertial frame of $E$. In so far as the equation (12) depends only on the relative velocity $V$ this equation is independent of the time of emission and independent of the choice of inertial frame. It is to be emphasised that the derivation of the standard formula (12) depends decisively on the use of the *Newtonian addition of velocities*. This fact is always masked in the textbook derivation of equation (3) by formulating the problem in the inertial *rest* frame of $E$ ( in which $c_0$ is replaced by $c$). The result (12) amply justifies our assumption of the Newtonian addition of velocities.

## II. $E$ and $R$ moving with constant acceleration $g$ in $S$

In this important particular case we deal with the experiment used in the *standard derivation* of the gravitational red shift formula (4c). $E$ and $R$ are now *fixed* in the *rigid* elevator with $E$ on the floor and $R$ on the ceiling of the elevator, directly above $E$. The elevator moves with constant acceleration $g$ along $\vec{ER}$, along the $x$-axis, say, of $S$. Since the elevator is *rigid*, in the classical sense, the distance between any two points of the elevator is constant; in particular,

$$x_R(t) - x_E(t) = \text{constant} = h, \text{say, for all } t. \quad (13)$$

Since

$$\ddot{x}_E(t) = g, \quad \ddot{x}_R(t) = g, \quad \text{for all } t, \quad (14a)$$

we get $\dot{x}_E(t) = gt + u$ and $\dot{x}_R(t) = gt + v$, $u$ and $v$ being the velocities of $E$ and $R$, respectively, at $t = 0$. But equation (13) implies that $\dot{x}_E(t) = \dot{x}_R(t)$ for all $t$. Hence

$u = v$ and, therefore,

$$\dot{x}_E(t) = gt + u = \dot{x}_R(t). \quad (14b)$$

Integrating equation (14b), and assuming that at $t = 0$ $x_E = a$, $x_R = a + h$, we get

$$x_E = \frac{1}{2}gt^2 + ut + a, \ x_R = \frac{1}{2}gt^2 + ut + a + h. \quad (14c)$$

Substituting equations (14) into equation (11) we get

$$[c_0 - g(T_1 - t_1)](\nu_R - \nu_E) = 0, \quad (15)$$

$c_0$ being the speed of light relative to $E$ at the instants of emission. The value of $T_1$ is obtained by substituting equations (14) into equation (8) and is given by the quadratic equation

$$gT_1^2 - 2(c_0 + gt_1)T_1 + (gt_1^2 + 2c_0t_1 + 2h) = 0. \quad (16)$$

For a given $t_1$ (and $g$ and $h$) this quadratic equation has *two* roots which are real only if $c_0^2 \geq 2gh$. Evidently the value of $T_1$ involved in equation (15) is the *smaller* root of equation (16) and is given, *exactly*, by

$$gT_1 = gt_1 + c_0 - c_0(1 - \frac{2gh}{c_0^2})^{\frac{1}{2}}. \quad (17)$$

[We add, parenthetically, that the existence of the second (larger) root of equation (16) is easy to understand. In classical mechanics there is nothing to prevent an object, e.g. the receiver $R$, attaining a speed greater than the speed of light. Since $R$ moves with constant acceleration $g$ it will eventually attain a speed greater than $c_0$ thus meeting the light ray $\Gamma_1$ at a second point (given by the larger root of equation (16))].

Returning to equation (15) we see that either

$$c_0 - g(T_1 - t_1) = 0 \qquad (18)$$

or

$$\nu_R - \nu_E = 0. \qquad (19)$$

The first possibility, together with equation (17), leads to the condition $c_0^2 = 2gh$. But $h$ and $g$ are completely *arbitrary* (even though, for definiteness, we took $g$ to be the acceleration due to gravity near the earth's surface). The first possibility must, therefore, be rejected. Thus in the present case (in which, let us recall, $E$ and $R$ are fixed in the constantly accelerating elevator), we must have, from equation (19),

$$\nu_R = \nu_E. \qquad (20)$$

This result is *exact* and it shows that *there is no Doppler shift* in this particular case. Hence, by the $EP$, *there is no gravitational red shift.* It follows, therefore, that the claim, made repeatedly for the last hundred years or so, that the $EP$ alone predicts the phenomenon of gravitational red shift, is *false*.

### III. $E$ at rest and $R$ moving with constant acceleration $g$ relative to $S$.

The main purpose of considering this particular case is to demonstrate that the problem dealt with in the *standard argument* is the problem in which $E$ is at *rest* and $R$ moves with *constant acceleration* relative to $S$; it is not, as is claimed in the standard argument, the problem in which both $E$ and $R$ partake in the constantly accelerating rigid elevator.

In this case $\dot{x}_E = 0$ and $\ddot{x}_R = g$ for all $t$. Assuming, only for simplicity, that at $t = 0$ $x_E = 0$, $x_R = h$ and $\dot{x}_R = 0$, we get

$$x_E = 0, \quad x_R = \frac{1}{2}gt^2 + h, \quad \dot{x}_R = gt \text{ for all } t \quad (21)$$

Substituting equation (21) in equation (11) we get

$$\nu_R = \left(1 - \frac{gT_1}{c}\right)\nu_E. \quad (22)$$

Here, by equation (21), $gT_1 = \dot{x}_R(T_1)$ is the velocity of $R$ relative to $E$, (and hence relative to $S$), at the time $t = T_1$ at which the first wave arrives at $R$, and $c$ is the speed of light in $S$. The actual value of $T_1$ can be found by substituting equation (21) into equation (8); it is given by the equation

$$gT_1^2 - 2cT_1 + 2(ct_1 + h) = 0 \quad (23)$$

For a given $t_1$ (and $g$ and $h$) this quadratic equation for $T_1$ has *two* roots which are real only if $2gct_1 < c^2 - 2gh$. Evidently the value of $T_1$ involved in equation (22) is the *smaller* root of equation (23); neglecting terms of order $c^{-2}$ this root is $T_1 = t_1 + h/c$. [The existence of the second (larger) root of equation (23) can be explained as is case $II$ above]. If, as in the standard argument, we take $t_1 = 0$, then $T_1 = h/c$ and equation (22) reduces to

$$\nu_R = \left(1 - \frac{gh}{c^2}\right)\nu_E. \quad (24)$$

This is *identical* to equation (4a) obtained by the standard argument. It is essential to recall that in the standard argument both $E$ and $R$ are supposed to be fixed in the constantly accelerating rigid elevator; that is, $E$ and $R$ move with the common acceleration $g$. The result obtained in this special case, case $III$, indicates conclusively that this is *not* the case. As already indicated in section 2, and as demonstrated in this special case, *the problem dealt with in the standard argument is the problem in which $E$ is at rest and $R$ moves with constant acceleration relative to the inertial frame $S$*.

## 4. Discussion

The reader is reminded that the gravitational red shift has been verified experimentally on the Earth's surface (Pound and Rebka[4], and Pound and Snider[5]; its existence is, therefore, beyond any doubt. Since, as we have shown in section 3, the $EP$ does not predict any red shift at all, the question naturally arises: *Is the equivalence principle wrong?* One possible answer is that the $EP$ is, indeed, *wrong*. This, however, is not the only answer. One may argue that the $EP$ is correct but that classical mechanics ( and, in particular, the assumptions of the existence of rigid bodies and of the validity of the Newtonian addition of velocities ) is unable to deal with electromagnetic or light phenomena. The only valid theory able to deal with such phenomena is the special theory of relativity. This is, of course, perfectly true. If we do take this point of view then, of course, the argument presented in this paper falls apart. But then so does the *standard argument* and classical mechanics cannot be used to investigate the *inertial* effects on the experiment in the accelerating elevator (case $II$).

It should be mentioned that Schild[6] has analysed the $EP$, as applied to case $II$, entirely within the framework of the special theory of relativity. There are, however, grave difficulties in such an analysis stemming from the facts that there are no rigid bodies (as ordinarily understood in classical mechanics) in special relativity and that *constant acceleration* is not an *invariant* concept in this theory. What *is* invariant in special relativity is the concept of *uniform acceleration* ( Bondi[16], Schild[6]). Without entering into any details, we mention that Schild[6] was able to construct an ingenious model of a "rigid" elevator using the concept of *uniform* acceleration, formulate case $II$ in such an elevator using special relativity and derive the standard gravitational red shift formula ($4c$). There is, however, a serious drawback in this analysis which makes Schild's arguments not sufficiently convincing. One cannot talk of *the* uniform acceleration of the "rigid" elevator model because different particles of the elevator, and in particular the emitter ($E$) and the receiver ($R$), move with *different* uniform accelerations. Schild was, of course, fully aware of the difference (variation) between the uniform accelerations of $E$ and $R$ but ignored it stating that "we can ignore this variation by making the elevator small,...". The red shift which he derives, however, is *precisely of the same order of magnitude as this variation*. Thus special relativity, too, is unable (so far, at any rate) to investigate the *inertial* effects on the experiment in case $II$.

It can be claimed, therefore, that the $EP$ is, indeed, *correct* but that the only two theories at our disposals, namely classical mechanics and special relativity, are unable to investigate the above inertial effects. This would, however, leave the $EP$ looking like mighty Hercules without his most important challenge ( to wit, the prediction of the gravitational red shift) to display his might.

The inescapable conclusion from the result in case $II$ is that either the $EP$ is wrong or, at the very least, it cannot predict the gravitational red shift which it was purported to do for so long. As was mentioned at the beginning, to the overwhelming majority of relativists this principle forms the cornerstone of the general theory of relativity. Wheeler and Ciufolini[17], in their book *Gravitation and Inertia* state: " At the foundation of Einstein's geometrodynamics (general relativity) and of its geometrical structure is one of the best-tested principles in the whole of physics: the equivalence principle." Indeed, many relativists go as far as saying that if the equivalence principle falls so does general relativity. Bergman[18], for example, says that "This (equivalence) principle cannot be eliminated without destroying the theory (of general relativity) as a whole."

In view of these widely held views, and in view of the failure of the $EP$ to predict the phenomenon of the gravitational red shift, the question must be asked: *Is the general theory of relativity wrong?* The answer is, of course, an emphatic and resounding *no*. As is shown in every book on general relativity the *full* theory (including its field equations), irrespective of its origins, predicts simply and unambiguously the exact experimentally observed gravitational red shift. Thus, by an ironic reversal of events, the non-validity of the $EP$, or at least its failure to predict the gravitational red shift which gave the principle so much prominence in the first place, not only does it not invalidate the general theory of relativity but it *strengthens* it; for the full general theory of relativity *alone* predicts the observed gravitational red shift.

It would be wrong to conclude this paper without mentioning that the equivalence principle has had, over the years, its fair share of criticism ( e.g., Fock[19], Synge[20] and, more recently, Ohanian and Ruffini[21]) and Ohanian[22]). Synge[20], for example, in his superb book on general relativity dismisses the $EP$ altogether. In his preface, the only place where he refers to the $EP$, he says: "... I have never been able to understand this principle. ... The principle of equivalence performed the essential office of midwife at the birth of general relativity. ... I suggest that the midwife be now buried with appropriate honours..."

After almost one hundred years since its birth, and having withstood with amazing success the most stringent experimental and observational tests, it can justifiably be claimed that the general theory of relativity has become of age. Irrespective of its *historical* development, the theory can stand unshakable on its own two feet. In view of the result obtained in this paper perhaps the time has come to follow, with very heavy heart indeed,

Synge's suggestion; namely, to bury the midwife at the birth of the theory with the highest possible honours.

The author is deeply grateful to Professor Brendan Scaife for his continued interest in this work, for long and critical discussions and for playing devil's advocate so convincingly. The author is also deeply indebted to the late Professor Dennis Sciama for a long and illuminating discussion on the principle of equivalence in general during his visit to Trinity College, Dublin, in 1998 to deliver the Fourth John Lighton Synge Public Lecture.# References